\documentclass{PoS}
\usepackage{bm,bbm,amssymb,graphicx}
\def\sl{\!\!\!\!\!/}
\newcommand{\tensor}[1]{\stackrel{\leftrightarrow}{#1}}
\newcommand{\tenD}{-\frac{i}{2}\!\!\stackrel{\leftrightarrow}{\bm{D}}}

\title{\vspace{-2cm}
\rightline{\normalsize\rm ANL-HEP-CP-13-3}
\vspace{1.3cm}
Order-$\bm{v^4}$ Relativistic Corrections to
\\ Gluon Fragmentation
into $\bm{{}^3S_1}$ Quarkonium}

\ShortTitle{Order-$v^4$ Relativistic Corrections to
Gluon Fragmentation
into ${}^3S_1$ Quarkonium}

\author{Geoffrey T. Bodwin\\
        HEP Division, Argonne National Laboratory\\
        E-mail: \email{gtb@hep.anl.gov}}

\author{U-Rae Kim\\
        Department of Physics, Korea University\\
        E-mail: \email{sadafada@korea.ac.kr}}

\author{\speaker{Jungil Lee}\\
        Department of Physics, Korea University\\
        E-mail: \email{jungil@korea.ac.kr}}

\abstract{
We compute the relativistic corrections to the color-singlet
contribution to gluon fragmentation into a $J/\psi$ at relative order
$v^4$, making use of the nonrelativistic QCD (NRQCD) factorization
approach. The corresponding full-QCD process exhibits infrared
divergences that manifest themselves as single and double poles in
$\epsilon$ in $4-2\epsilon$ dimensions. We isolate the
infrared-divergent contributions and treat them analytically. In the
matching of full QCD to NRQCD, the pole contributions are absorbed into
long-distance NRQCD matrix elements. The renormalizations of the
ultraviolet divergences of the long-distance NRQCD matrix elements
involve Born and one-loop single-pole counterterm contributions and Born
double-pole counterterm contributions. While the order-$v^4$
contribution enhances the $J/\psi$ hadroproduction rate for the
color-singlet channel substantially, this contribution is not
important numerically in comparison with the color-octet contributions.
We also find that the ${}^3P_J$ color-octet channel in the gluon
fragmentation function contributes to $J/\psi$ hadroproduction
significantly in comparison with the complete contribution of
next-to-leading order in $\alpha_s$ in that channel. }

\FullConference{Xth Quark Confinement and the Hadron Spectrum\\
                 8-12 October 2012\\
                 TUM Campus Garching, Munich, Germany}
\begin{document} 

\section{Introduction}
In this proceedings contribution, we summarize the computation of the
fragmentation of a gluon into a spin-triplet $S$-wave quarkonium via the
${}^3S_1^{[1]}$ channel in relative order $v^4$, where $v$ is the
velocity of the heavy quark $Q$ or heavy antiquark $\bar Q$ in the
quarkonium rest frame. We refer the reader to Ref.~\cite{Bodwin:2012xc}
for details of this calculation. The expression ${}^{2s+1}l_j^{[c]}$
is the standard spectroscopic notation for the spin $s$, orbital angular
momentum $l$, total angular momentum $j$, and color $c=1$ (singlet) or 8
(octet) of the $Q\bar{Q}$ pair that is created at short distances and
that evolves into the heavy quarkonium.

Our computation make use of the nonrelativistic QCD (NRQCD)
factorization approach \cite{BBL}. That is, we express the
fragmentation functions as sums of products of NRQCD long-distance
matrix elements (LDMEs) and short-distance coefficients. The focus of
our calculation is the short-distance coefficient that appears in the
order $v^4$ contribution to the fragmentation function for the
${}^3S_1^{[1]}$ channel.  The short-distance coefficients for gluon
fragmentation in this channel in relative orders $v^0$ and $v^2$ have
already been computed in Refs.~\cite{Braaten:1993rw,Braaten:1995cj} and
\cite{Bodwin:2003wh}, respectively. The contribution of relative order
$v^4$ is interesting theoretically because it is at this order that the
${}^3S_1^{[1]}$ fragmentation channel first develops soft divergences.
These soft divergences are ultimately absorbed into the NRQCD LDME
for the evolution of a ${}^3S_1^{[8]}$ or ${}^3P_J^{[8]}$ $Q\bar Q$ pair
into a spin-triplet $S$-wave quarkonium state. Consequently, the
${}^3S_1^{[1]}$ contribution in order $v^4$ and the ${}^3S_1^{[8]}$ and
${}^3P_J^{[8]}$ contributions at the leading nontrivial order in $v$
are related by logarithms of the factorization scale. Since the
${}^3S_1^{[8]}$ contribution to the fragmentation of a gluon into a
$J/\psi$ is known to be significant phenomenologically, it is important
to compute the ${}^3S_1^{[1]}$ contribution in order $v^4$, as well.

An important technical issue in this calculation is the appearance of
both single and double soft poles in the dimensional-regularization 
parameter $\epsilon$. This is the first NRQCD factorization
calculation in which double soft poles have appeared. In order to effect
the matching between NRQCD and full QCD that removes these poles, it is
necessary to work out, for the first time, two-loop corrections 
to NRQCD operators and the corresponding ultraviolet 
renormalizations.

We employ the Collins-Soper definition \cite{Collins:1981uw} of the
fragmentation function for a gluon fragmenting into a quarkonium. 
We compute the full-QCD fragmentation functions
for a gluon fragmenting into free $Q\bar Q$ states with various quantum
numbers. We then determine the NRQCD short-distance coefficients by
comparing the full-QCD fragmentation functions with the corresponding
NRQCD expressions, making use of the NRQCD factorization formulas.

The remainder of this paper is organized as follows. In
Sec.~\ref{sec:FF}, we present the NRQCD factorization formulas through
relative order $v^4$ for the fragmentation functions for a gluon
fragmenting into a $J/\psi$. The results of the full-QCD calculation
for the corresponding fragmentation functions for free $Q\bar{Q}$ states
are given in Sec.~\ref{sec:full-qcd}. In Sec.~\ref{sec:ldmes}, we
compute the relevant NRQCD LDMEs for free $Q\bar{Q}$ states analytically
in dimensional regularization, and we determine the evolution
equations for the LDMEs. In Sec.~\ref{sec:short-d}, we compute the
short-distance coefficients by matching the NRQCD and full-QCD results.
Sec.~\ref{sec:numerical} contains estimates of the relative sizes of the 
various fragmentation contributions to the $J/\psi$ hadroproduction cross 
section. We give a summary of our results in
Sec.~\ref{sec:discussion}.

\section{Factorization formulas}
\label{sec:FF}%
We denote by $D[g \to {J/\psi}](z,\mu_\Lambda)$ the fragmentation
function for a gluon fragmenting into a $J/\psi$. Here, $\mu_\Lambda$
is the factorization scale, $z=P^+/k^+$, and $P$ and $k$ are the
momenta of the ${J/\psi}$ and the fragmenting gluon, respectively.
We define light-cone coordinates for a four-vector
$V=(V^+,V^-,\bm{V}_\bot)$ by $V^\pm \equiv(V^0\pm V^3)/\sqrt{2}$. We
make use of the Collins-Soper definition of $D[g \to
{J/\psi}](z,\mu_\Lambda)$ \cite{Collins:1981uw}, which is gauge
invariant, and we regularize soft and ultraviolet divergences
dimensionally, taking $d=4-2\epsilon$ space-time dimensions.

The NRQCD factorization formula for $D[g \to {J/\psi}](z,\mu_\Lambda)$
is given by
\begin{equation}
D[g \to J/\psi](z)=
\sum_{n} 
d_{n}(z) \langle 0| \mathcal{O}_{n}^{J/\psi} |0\rangle,
\label{eq:NRQCD-fact-frag}
\end{equation}
where the $\langle 0|\mathcal{O}_{n}^{J/\psi} |0\rangle$ are NRQCD LDMEs,
the $d_{n}(z)$ are short-distance coefficients, and we have
suppressed the $\mu_\Lambda$ dependences in $d_{n}(z)$ and  $\langle
0|\mathcal{O}_{n}^{J/\psi} |0\rangle$.
Since the $d_{n}(z)$ are independent of the hadronic final
state, we can write the fragmentation function for a gluon fragmenting
into a free $Q\bar Q$ state as
\begin{equation}
D[g \to Q\bar Q](z)=
\sum_{n} 
d_{n}(z) \langle 0| \mathcal{O}_{n}^{Q\bar Q} |0\rangle.
\label{eq:NRQCD-fact-frag-free}
\end{equation}
By computing the left side of Eq.~(\ref{eq:NRQCD-fact-frag-free}) in
full QCD and comparing it with the right side, we can determine the
$d_{n}(z)$. In the remainder of this paper, we denote the order-$v^k$
contribution to $D[g\to H]$ by $D_k[g\to H]$, where $H$ can be either a
quarkonium or a free-$Q\bar Q$ final state.

The operator LDMEs that we use in this paper are
\begin{eqnarray}
\langle 0|\mathcal{O}_{0}^H({}^3S_1^{[1]})|0\rangle
&=&
\langle 0|\chi^\dagger \sigma^i\psi 
\; {\cal P}_{H} \;
\psi^\dagger \sigma^i \chi|0\rangle,
\\
\langle 0|\mathcal{O}_{2}^H({}^3S_1^{[1]}) |0\rangle 
&=&
\frac{1}{2}\langle 0|\chi^\dagger \sigma^i 
(\tenD)^2\psi \; 
{\cal P}_{H} \;
        \psi^\dagger \sigma^i \chi +\textrm{H.~c.} |0\rangle,
\\
\langle 0|{\cal O}_{0}^H({}^1S_0^{[8]})|0\rangle
&=&\langle 0|\chi^\dagger T^a\psi\;{\cal P}_{H}\;
\psi^\dagger T^a \chi |0\rangle,
\\
\langle 0|{\cal O}_{4,1}^H({}^3S_1^{[1]})|0\rangle
&=&                                   
\langle 0|
\chi^\dagger \sigma^i
(\tenD)^2                                                 
\psi
\;{\cal P}_{H}\;
\psi^\dagger \sigma^i
(\tenD)^2 
\chi
|0\rangle,
\\
\langle 0|\mathcal{O}_{4,2}^H({}^3S_1^{[1]}) |0\rangle &=&
\frac{1}{2}\langle 0|\chi^\dagger \sigma^i 
(\tenD)^4
\psi \; 
{\cal P}_{H} \;
        \psi^\dagger \sigma^i \chi +\textrm{H.~c.} |0\rangle,\\
\langle 0|\mathcal{O}_{4,3}^H({}^3S_1^{[1]}) |0\rangle &=&
\frac{1}{2}\langle 0|\chi^\dagger \sigma^i \psi
\;{\cal P}_{H} \;
\psi^\dagger \sigma^i 
(\tensor{\bm{D}}\!\cdot g_s \bm{E}+g_s\bm{E}\cdot\!\tensor{\bm{D}})
\chi\nonumber\\
&&\qquad -\chi^\dagger \sigma^i 
(\tensor{\bm{D}}\!\cdot g_s \bm{E}+g_s\bm{E}\cdot\!\tensor{\bm{D}})\psi
\;{\cal P}_{H} \;
\psi^\dagger \sigma^i \chi |0\rangle,\\
\langle 0| {\cal O}_{0}^H({}^3S_1^{[8]})|0\rangle
&=&                              
\langle 0|
\chi^\dagger \sigma^i T^a
\psi
\;{\cal P}_{H}\;                     
\psi^\dagger \sigma^i T^a
\chi|0\rangle,
\\
\langle 0| {\cal O}_{0}^H({}^3P^{[8]})|0\rangle
&=&\langle 0|                                      
\chi^\dagger      
(-\frac{i}{2}\!\!\stackrel{\leftrightarrow}{\bm{D}}
)^r \sigma^n T^a
\psi
\;{\cal P}_{H}\;
\psi^\dagger
(-\frac{i}{2}\!\!\stackrel{\leftrightarrow}{\bm{D}})^r \sigma^n T^a
\chi
|0\rangle,
\end{eqnarray}
where $g_s=\sqrt{4\pi\alpha_s}$, $\psi^\dagger$ and $\chi$ are
two-component (Pauli) fields that create a heavy quark and a heavy
antiquark, $\bm{D}$ is the gauge-covariant derivative, and $\bm{E}$ is
the chromoelectric field operator. ${\cal
P}_{H(P)}=\sum_X|H(P)+X\rangle\langle H(P)+X|$ is a projection onto a
state consisting of a quarkonium $H$, with four-momentum $P$, plus
anything. ${\cal P}_{H(P)}$ contains a sum over any quarkonium
polarization quantum numbers that are not specified explicitly. The
symmetric traceless product is defined by $A^{(i}B^{j)}=\frac{1}{2}(A^i
B^j+A^j B^i) -\frac{1}{d-1}\,\delta^{ij}A^k B^k$, and the antisymmetric
product is defined by $ A^{[i}B^{j]}=\frac{1}{2}(A^iB^j-A^jB^i)$.

We now give the NRQCD factorization formulas for gluon fragmentation
into a $J/\psi$ through relative order $v^4$. The contributions to $D[g
\to J/\psi]$ in relative orders $v^0$, $v^2$, and $v^3$ are
\begin{eqnarray}
\label{D0-J/psi}
D_0[g \to J/\psi]
&=&
d_0[g \to Q\bar{Q}({}^3S_1^{[1]})]
\langle 0|
\mathcal{O}_{0}^{J/\psi}({}^3S_1^{[1]})
|0\rangle,
\\
\label{D2-J/psi}
D_2[g \to J/\psi]
&=&
d_2[g \to Q\bar{Q}({}^3S_{1}^{[1]})]
\langle 0|
\mathcal{O}_{2}^{J/\psi}({}^3S^{[1]})
|0\rangle,
\\
\label{D3-J/psi}
D_3[g \to J/\psi]
&=&
d_0[g \to Q\bar{Q}({}^1S_0^{[8]})]
\langle 0|
\mathcal{O}_{0}^{J/\psi}({}^1S_0^{[8]})
|0\rangle.
\end{eqnarray}
The short-distance coefficients $d_0[g \to Q\bar{Q}({}^3S_1^{[1]})]$ and
$d_2[g \to Q\bar{Q}({}^3S_1^{[1]})]$ were calculated previously in
Refs.~\cite{Braaten:1993rw,Braaten:1995cj} and
Ref.~\cite{Bodwin:2003wh}, respectively. The short-distance coefficient
$d_0[g \to Q\bar{Q}({}^1S_0^{[8]})]$ is related by an overall color
factor to $d_0[g \to Q\bar{Q}({}^1S_0^{[1]})]$, which was calculated in
Ref.~\cite{Braaten:1996rp}. The contribution to $D[g \to J/\psi]$ in
relative order $v^4$ is
\begin{eqnarray}
\label{D4-Jpsi}
D_{4}[g \to J/\psi]
&=&
\big\{\,
d_{4,1}[g \to Q\bar{Q}({}^3S_1^{[1]})]
+
d_{4,2}[g \to 
Q\bar{Q}({}^3S_1^{[1]})]\,\big\}
\,
\langle 0|
\mathcal{O}_{4}^{J/\psi}({}^3S_1^{[1]})
|0\rangle
\nonumber\\
&+&
d_{0}[g \to Q\bar{Q}({}^3P^{[8]})]
\langle 0|
\mathcal{O}_{0}^{J/\psi}({}^3P^{[8]})
|0\rangle
\nonumber\\
&+&
d_{0}[g \to Q\bar{Q}({}^3S_1^{[8]})]
\langle 0|
\mathcal{O}_{0}^{J/\psi}({}^3S_1^{[8]})
|0\rangle.
\end{eqnarray}
The short-distance coefficient $d_{0}[g \to Q\bar{Q}({}^3S_1^{[8]})]$
was calculated previously in
Refs.~\cite{Bodwin:2003wh,Braaten:1996rp,BL:gfrag-NLO,Lee:2005jw}. The
short-distance coefficient $d_{0}[g \to Q\bar{Q}({}^3P^{[8]})]$ is
related by an overall color factor to the sum over $J$ of the
short-distance coefficients $d_{0}[g \to Q\bar{Q}({}^3P_J^{[1]})]$ that
were calculated in Ref.~\cite{Braaten:1996rp}. The computation of the
combination of short-distance coefficients $d_{4,1}[g \to
Q\bar{Q}({}^3S_1^{[1]})]+d_{4,2}[g \to Q\bar{Q}({}^3S_1^{[1]})]$ is the
main goal of this work. \footnote{In fact, the color-singlet
contribution to $D_{4}[g \to J/\psi]$ is a linear combination of the
LDMEs $\langle 0|\mathcal{O}_{4,i}^{J/\psi}({}^3S_1^{[1]})
|0\rangle$, for $i=1$--$3$. We express the contribution of $\langle
0|\mathcal{O}_{4,3}^{J/\psi}({}^3S_1^{[1]}) |0\rangle$ in terms of the
other two LDMEs by making use of the NRQCD equations of 
motion, and we use the vacuum-saturation approximation to replace 
the $\langle 0|\mathcal{O}_{4,i}^{J/\psi}({}^3S_1^{[1]}) |0\rangle$
with a single
LDME, $\langle 0|\mathcal{O}_{4}^{J/\psi}({}^3S_1^{[1]}) |0\rangle$,
making an error of relative order $v^2$ \cite{Bodwin:2002hg}.}

\section{Full-QCD calculations\label{sec:full-qcd}} 

We compute the fragmentation functions for free $Q\bar{Q}$ states in
full QCD in $d=4-2\epsilon$ dimensions, multiplying the square of the
dimensional-regularization scale $\mu$ by the factor
$e^{\gamma_{\rm\,E}}/(4\pi)$ that is appropriate to the
modified-minimal-subtraction ($\overline{\rm MS}$) scheme. Here,
${{\gamma}_{}}_{\rm E}$ is the Euler-Mascheroni constant.

The results for $D_0[g\to Q\bar Q({}^3S_1^{[8]})]$ and $D_{2}[g\to
Q\bar Q({}^3P^{[8]})]$ at leading order (LO) in $\alpha_s$ and $v$ are
\begin{eqnarray}
D_0[g\to Q\bar Q({}^3S_1^{[8]})]
&=&
\frac{\pi\alpha_s}{m^3}\left(
\frac{\mu^2}{4\pi}e^{{{\gamma}_{}}_{\rm E}}
\right)^\epsilon\delta(1-z),
\\
D_{2}[g\to Q\bar Q({}^3P^{[8]})]&=&
\frac{8\alpha_s^2\,\bm{q}^2}{(d-1)m^5}
\,
\frac{N_c^2-4}{4N_c}
(1-\epsilon)
\Gamma(1+\epsilon)
\left(
\frac{\mu^2}{4\pi}e^{{{\gamma}_{}}_{\rm E}}
\right)^{\epsilon}
\left(\frac{\mu^2}{4m^2}e^{{{\gamma}_{}}_{\rm E}}\right)^\epsilon
\nonumber\\
&&\times
\left[-\frac{1}{2\epsilon_{\rm IR}}\, \delta(1-z) 
+  f(z)\right],
\end{eqnarray}
where  $N_c$ is the number of colors, $\bm{q}$ is half the relative
momentum of the $Q$ and $\bar{Q}$ in the $Q\bar{Q}$ rest frame, and the
finite function $f(z)$ is defined in Eq.~(5.20) of
Ref.\cite{Bodwin:2012xc}. The subscript ``IR'' in $\epsilon_{\rm
IR}^{-1}$ indicates that the pole is infrared in origin.

The result for $D_{4}[g\to Q\bar Q({}^3S_1^{[1]})]$
at LO in $\alpha_s$ is
\begin{equation}
\label{D4-tot}
D_4[g\to Q\bar Q({}^3S_1^{[1]})]=
D_4[g\to Q\bar Q({}^3S_1^{[1]})]^{\rm 
finite}+ I[S_{12}]+I[S_1]+I[S_2],
\end{equation}
where  $I[S_{12}]$, $I[S_1]$, and $I[S_2]$ contain the double- 
and single-pole contributions: 
\begin{eqnarray}      
\label{5-i12-ans}       
I[S_{12}]
&=&
\left\{\frac{1}{8 \epsilon_{\rm IR}^2}\,\delta(1-z)
-\frac{1}{2 \epsilon_{\rm IR}}\left[\frac{1}{(1-z)^{1+4\epsilon}}\right]_{+}
+\frac{1-z^{1+2\epsilon}}{2 \epsilon_{\rm IR}(1-z)^{1+4\epsilon}}
\right\}\nonumber\\
&&\times
\left(
\frac{8\alpha_s}{3\pi m^{2}}
\right)^2
\frac{N_c^2-4}{16N_c^2}
\,
\frac{\pi\alpha_s}{(d-1) m^3}
\left(\frac{\mu^2}{4\pi}e^{{{\gamma}_{}}_{\rm E}}\right)^{\epsilon}
\frac{\bm{q}^4}{d-1}
\nonumber\\
&&\times
\left(\frac{\mu^2}{4m^2}e^{{{\gamma}_{}}_{\rm E}}\right)^{2\epsilon}
\frac{\Gamma^{2}(1+\epsilon)\Gamma^{2}(1-2\epsilon)}
{\Gamma(1-4\epsilon)}
(1-\epsilon)(6-2\epsilon-\epsilon^2-2\epsilon^3)
,
\\
I[S_{1}]&=&I[S_{2}]=
\left(-
\frac{\tau_1}{2\epsilon_{\rm IR}}+\tau_0
\right)
\left(\frac{\mu^2}{4m^2}e^{{{\gamma}_{}}_{\rm E}}\right)^{2\epsilon}
\frac{z^{-2+2\epsilon}(1-z)^{-4\epsilon}\Gamma^2(1+\epsilon)}
{48(1-\epsilon)}
\nonumber\\
&\times&
\left(
\frac{8\alpha_s}{3\pi m^2}
\right)^2
\frac{N_c^2-4}{16N_c^2}
\left(\frac{\mu^2}{4\pi}e^{{{\gamma}_{}}_{\rm E}}\right)^{\epsilon}
\frac{\pi\alpha_s}{(d-1)m^3}
\frac{\bm{q}^4}{d-1}+O(\epsilon).
\end{eqnarray}
Here, $\tau_1$ and $\tau_0$ are functions of $z$ that are defined in
Eq.~(5.29) of Ref.\cite{Bodwin:2012xc}. $D_4[g\to Q\bar
Q({}^3S_1^{[1]})]^{\rm finite}$ in Eq.~(\ref{D4-tot}) is a finite
contribution, which we evaluate numerically in $d=4$ dimensions.

\section{NRQCD LDMEs\label{sec:ldmes}}
In this section we tabulate our results for the NRQCD LDMEs for free
$Q\bar Q({}^3S_1^{[1]})$ states that are relevant through relative order
$v^4$.

The free $Q\bar{Q}$ matrix elements at order $\alpha_s^0$ are normalized as 
\begin{eqnarray}
\label{eq:normalization}
\langle 0|
\mathcal{O}_{0}^{Q\bar{Q}({}^1S_0^{[8]})}({}^1S_0^{[8]})
|0\rangle^{(0)}
&=&(N_c^2-1),\\
\langle 0|\mathcal{O}_{0}^{Q\bar{Q}({}^3S_1^{[1]})}({}^3S_1^{[1]})
|0\rangle^{(0)}
&=&2(d-1)N_c,
\label{eq:normalization-a}
\\
\langle 0|\mathcal{O}_{0}^{Q\bar{Q}({}^3S_1^{[8]})}({}^3S_1^{[8]})
|0\rangle^{(0)}
&=&(d-1)(N_c^2-1),
\label{norm-3s18}
\\
\langle 0|
\mathcal{O}_{2}^{Q\bar{Q}({}^3S_1^{[n]})}({}^3S_1^{[n]})
|0\rangle^{(0)}
&=&\bm{q}^2
\langle 0|
\mathcal{O}_{0}^{Q\bar{Q}({}^3S_1^{[n]})}({}^3S_1^{[n]})
|0\rangle^{(0)},
\label{eq:normalization-c}
\\
\label{eq:normalization-3s1-4}
\langle 0|\mathcal{O}_{4}^{Q\bar{Q}({}^3S_1^{[n]})}({}^3S_1^{[n]})
|0\rangle^{(0)}
&=&\bm{q}^4
\langle 0|
\mathcal{O}_{0}^{Q\bar{Q}({}^3S_1^{[n]})}({}^3S_1^{[n]})
|0\rangle^{(0)},\\
\label{eq:normalization-3p8}
\langle 0|
\mathcal{O}_{0}^{Q\bar{Q}({}^3P^{[8]})}({}^3P^{[8]})
|0\rangle^{(0)}
&=& \bm{q}^2(d-1)(N_c^2-1),
\end{eqnarray}
where there is an implied sum over final-state polarizations,
and the superscript $(k)$ indicates the order in $\alpha_s$.

In order $\alpha_s$ and order $\alpha_s^2$, the relevant LDMEs 
mix:
\begin{eqnarray}                                                     
\langle 0| {\cal O}_{0}^{Q\bar 
Q({}^3S_1^{[1]})}({}^3P^{[8]})|0\rangle_{\overline{\rm MS}}^{(1)}
&=&
\frac{8\alpha_s}{3\pi m^2}
\left(\frac{-1}{2\epsilon_{\rm IR}}\right)
\frac{N_c^2-1}{4N_c^2}
\,
\frac{1}{d-1}
\langle 0| {\cal O}_{4,1}^{Q\bar 
Q({}^3S_1^{[1]})}({}^3S_1^{[1]})|0\rangle^{(0)},
\label{3pj3s1msbar}
\\
\langle 0| {\cal O}_{0}^{Q\bar 
Q({}^3P^{[8]})}({}^3S_1^{[8]})|0\rangle_{\overline{\rm MS}}^{(1)}
&=&\frac{8\alpha_s}{3\pi m^2}
\left(\frac{-1}{2\epsilon_{\rm IR}}\right)
\frac{N_c^2-4}{4N_c}
\langle 0| {\cal O}_{0}^{Q\bar 
Q({}^3P^{[8]})}({}^3P^{[8]})|0\rangle^{(0)},\phantom{xxx}
\label{3s13pjmsbar}
\\
\langle 0| {\cal O}_{0}^{Q\bar 
Q({}^3S_1^{[1]})}({}^3S_1^{[8]})|0\rangle_{\overline{\rm MS}}^{(2)}
&=&
\frac{1}{8\epsilon_{\rm IR}^2}
\left(
\frac{8\alpha_s}{3\pi m^2}\right)^2
\frac{(N_c^2-1)(N_c^2-4)}{16N_c^3}
\frac{1}{d-1}\langle 0| {\cal O}_{4,1}^{Q\bar 
Q({}^3S_1^{[1]})}({}^3S_1^{[1]})|0\rangle^{(0)}.
\nonumber\\
\label{3s13s1msbar}
\end{eqnarray} 
Here, the subscript $\overline{\rm MS}$ indicates that we have 
removed the ultraviolet poles in $\varepsilon$ by using the
minimal-subtraction procedure, with the choice of scale that is
appropriate to $\overline{\rm MS}$ subtraction. The renormalization
subtractions for $\langle 0| {\cal O}_{0}^{Q\bar
Q({}^3S_1^{[1]})}({}^3S_1^{[8]})|0\rangle^{(2)}$ involve both Born
diagrams for the double-pole counterterm and one-loop diagrams for the
single-pole counterterm. In the minimal subtraction procedure, one
subtracts only pure pole contributions in the ultraviolet-divergent 
subdiagrams. It is essential for the consistency of the
minimal-subtraction program to treat factors that are external to those
divergent subdiagrams, such as angular-momentum projections
involving external momenta, exactly in $d=4-2\varepsilon$ dimensions. A
failure to follow this procedure in the presence of poles in
$d=4-2\varepsilon$ of order two or higher can result in the appearance
of uncanceled poles in $\varepsilon$ in the NRQCD short-distance
coefficients. We refer the reader to Ref.~\cite{Bodwin:2012xc} for
details.

The dimensional-regularization scale $\mu$ can be identified
with the NRQCD factorization scale $\mu_\Lambda$. It then follows from 
$d\alpha_s/d\log(\mu_\Lambda)=-2\varepsilon \alpha_s+O(\alpha_s^2)$ that 
the renormalization-group evolution equations for the LDMEs are
\begin{eqnarray}
\frac{d}{d\log \mu_{\Lambda}}\langle 0| {\cal O}_{0}^{Q\bar Q({}^3S_1^{[1]})}
({}^3P^{[8]})|0\rangle_{\overline{\rm MS}}^{(1)}
&=&
\frac{8\alpha_s}{3\pi m^2}\,
\frac{N_c^2-1}{4N_c^2}\,
\frac{1}{d-1}
\langle 0| {\cal O}_{4,1}^{Q\bar 
Q({}^3S_1^{[1]})}({}^3S_1^{[1]})|0\rangle^{(0)},
\label{3pj3s1-evo}
\\
\frac{d}{d\log \mu_{\Lambda}}
\langle 0| {\cal O}_{0}^{Q\bar Q({}^3P^{[8]})}
({}^3S_1^{[8]})|0\rangle_{\overline{\rm MS}}^{(1)}
&=&\frac{8\alpha_s}{3\pi m^2}
\frac{N_c^2-4}{4N_c}
\langle 0| {\cal O}_{0}^{Q\bar 
Q({}^3P^{[8]})}({}^3P^{[8]})|0\rangle^{(0)},
\phantom{xxx}
\label{3s13pj-evo}
\\
\frac{d}{d\log \mu_{\Lambda}}\langle 0| {\cal O}_{0}^{Q\bar Q({}^3S_1^{[1]})}
({}^3S_{1}^{[8]})|0\rangle_{\overline{\rm MS}}^{(2)}
&=&
\frac{8\alpha_s}{3\pi m^2}\,
\frac{N_c^2-4}{4N_c}
\langle 0| {\cal O}_{0}^{Q\bar Q({}^3S_1^{[1]})}
({}^3P^{[8]})|0\rangle_{\overline{\rm MS}}^{(1)}.
\label{3s13s1-evo2}
\end{eqnarray}
The result in Eq.~(\ref{3pj3s1-evo}) agrees with the corresponding
results of Refs.~\cite{Gremm:1997dq,z-g-he}.
Equations~(\ref{3s13pj-evo}) and (\ref{3s13s1-evo2}) agree with the result
in Eq.~(B19b) of Ref.~\cite{BBL} at the leading nontrivial order in $v$
and with the corresponding result in Ref~\cite{z-g-he}, but disagree
with the corresponding result in Ref.~\cite{Gremm:1997dq}.
\section{Short-distance coefficients\label{sec:short-d}}
\begin{figure}
\begin{center}
\includegraphics[width=0.4\columnwidth]{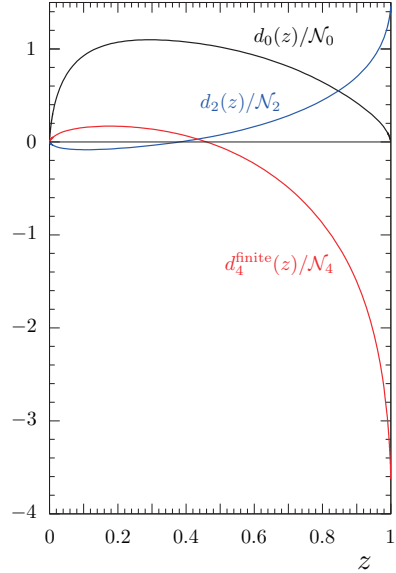}
\end{center}
\vspace{-.5cm}
\caption[]{The color-singlet short-distance coefficients
$d_0[g \to Q\bar{Q}({}^3S_1^{[1]})]^{(3)}(z)$,
$d_2[g \to Q\bar{Q}({}^3S_1^{[1]})]^{(3)}(z)$ and
$d_4[g \to Q\bar{Q}({}^3S_1^{[1]})]^{\rm finite}(z)$,
which are defined in Eqs.~(\ref{D0-J/psi}),
(\ref{D2-J/psi}) and (\ref{d412sum-ans}), respectively,
as functions of $z$.
The scaling factors  are 
$(\mathcal{N}_0,\mathcal{N}_2,\mathcal{N}_4)
=(10^{-3}\times\alpha_s^3/m^{3},
10^{-2}\times\alpha_s^3/m^{5},
10^{-2}\times\alpha_s^3/m^{7})$.
\label{fig:numerical}}
\end{figure}

By making use of the results of Secs.~\ref{sec:full-qcd} and
\ref{sec:ldmes} and the free-$Q\bar Q$ versions of the NRQCD
factorization (matching) equations (\ref{D0-J/psi}), (\ref{D2-J/psi}),
and (\ref{D4-Jpsi}) in Sec.~\ref{sec:FF}, we obtain
\begin{eqnarray}
\label{d412sum-ans}
&&
d_{4,1}[g \to Q\bar{Q}({}^3S_1^{[1]})]^{(3)}
+d_{4,2}[g \to Q\bar{Q}({}^3S_1^{[1]})]^{(3)}
\nonumber\\
&&\qquad=\,\,
d_{4}[g \to Q\bar{Q}({}^3S_1^{[1]})]^{\rm finite}
\,\,
+\,\,\frac{2\alpha_s^3 (N_c^2 -4)}{3\pi (d - 1)^3 N_c^3  m^7 }
\,\,\bigg\{\,\,
\delta(1-z)\,
\bigg(
\frac{1}{24}
-\frac{\pi^2}{6}
\nonumber\\
&&\phantom{xxxxx}
-\frac{1}{3}\log\frac{\mu_{\Lambda}}{2m}
+\log^2\frac{\mu_{\Lambda}}{2m}
\bigg)
+\left(
\frac{1}{1-z}
\right)_{\!\!+}\!\!
\bigg(
\frac{1}{3}-2\log\frac{\mu_{\Lambda}}{2m}
\bigg)
+2\left[
\frac{\log(1-z)}{1-z}
\right]_{\!+}\!
\nonumber\\
&&\phantom{xxxxx}
-\frac{104  - 29 z - 10 z^2 }{24}
+\frac{7[z+(1+z)\log(1-z)]}{2z^2}   
   + \frac{(1-2 z) (8-5 z) }{4}\log\frac{\mu_{\Lambda}}{2m}
\nonumber\\
&&\phantom{xxxxx}
   + \frac{1 + z}{4} \left(31 - 6 z - \frac{36}{z}\right) 
        \log(1 - z)
   - \frac{z}{4}  \left(\! 39 - 6 z +  \frac{8}{1-z}
                       \!\right) \log z
\nonumber\\
&&\phantom{xxxxx}
   + \frac{13-7z}{2}\bigg[\,
              \bigg(\!
                 \log \frac{1-z}{z^2}- \log\frac{\mu_{\Lambda}}{2m}
                 \bigg) \log(1-z)   -\textrm{Li}_2(z)\,
             \bigg]\,\,   
\bigg\}+O(\epsilon).
\end{eqnarray}
Here, $d_4[g\to Q\bar Q({}^3S_1^{[1]})]^{\rm finite}= {D_4[g\to
Q\bar Q({}^3S_1^{[1]})]^{\rm finite}}/ {\langle
0|\mathcal{O}_{4}^{Q\bar{Q}({}^3S_1^{[1]})}({}^3S_1^{[1]})
|0\rangle^{(0)}}$ is evaluated by carrying out the integrations
over the phase space numerically. The results of the numerical
integration are shown in Fig. \ref{fig:numerical}. One can find
details of the matching procedure in Sec.~7 of
Ref.~\cite{Bodwin:2012xc}.
\section{Relative sizes of the fragmentation contributions
\label{sec:numerical}}
We now make estimates of the relative sizes of the contributions of the
various fragmentation functions to the cross section for $J/\psi$
production in hadron-hadron collisions. At large $p_T$, the differential
cross section for a gluon can be approximated as
${d\sigma_{g}}/{d{{p}_{}}_{T}}\propto 1/{{p}_{}}_{T}^\kappa$, where
$\kappa$ is a fixed power. Then, the fragmentation contribution to the
$J/\psi$ production rate becomes $d\sigma_{J/\psi}^{\rm
frag}/d{{p}_{}}_{T}\propto I_{\kappa}(D)$, where $I_{\kappa}(D)=\int_0^1
\! dz \, z^{\kappa}\, D(z)$ \cite{Bodwin:2012xc}.
Hence, we can obtain a rough estimate of the relative contribution of a
fragmentation process to the cross section by computing $I_{\kappa}(D)$.
At large $p_{_T}$, the cross section at LO in $\alpha_s$ is dominated by
fragmentation of a gluon into a $J/\psi$ with $z=1$ \cite{Cho:1995vh}.
The next-to-leading order (NLO) $k$~factor for this channel is
essentially independent of $p_{_T}$ for the cross sections that were
measured at the Tevatron \cite{Gong:2008ft}. Making use of these facts
and the result for the ${}^3S_1^{[8]}$ contribution to $d\sigma/dp_{_T}$
at NLO in $\alpha_s$ that appears in Fig.~1(c) of
Ref.~\cite{Butenschoen:2010rq}, we have determined that $\kappa\approx
5.2$.

We can compare the relative contributions to $d\sigma_{J/\psi}^{\rm
frag}$ of each of the three $Q\bar Q$ channels that contribute
to $D[g\to J/\psi]$ in order $v^4$ [Eq.~(\ref{D4-Jpsi})]. For each
channel, we compute $I_{5.2}(D)$. Then, we multiply by the following
LDMEs:
$\langle 0|\mathcal{O}_{4}^{J/\psi}({}^3S_1^{[1]})
|0\rangle=m^4\langle v^2\rangle^2 \times 1.32~{\rm GeV}^3$,
$\langle 0|
\mathcal{O}_{0}^{J/\psi}({}^3P^{[8]})
|0\rangle=-0.109~{\rm GeV}^5$,
$\langle 0|
\mathcal{O}_{0}^{J/\psi}({}^3S_1^{[8]})
|0\rangle=3.12\times 10^{-3}~{\rm GeV}^3$,
$\langle 0|
\mathcal{O}_{0}^{J/\psi}({}^1S_0^{[8]})
|0\rangle=4.50\times10^{-2}~{\rm GeV}^3$.
We have obtained these LDMEs from Ref.~\cite{Butenschoen:2010rq}. In
the case of the first LDME, we have used the generalized Gremm-Kapustin
relation \cite{Bodwin:2007fz} $\langle
0|\mathcal{O}_{4}^{J/\psi}({}^3S_1^{[1]}) |0\rangle=m^4\langle
v^2\rangle^2 \langle 0|\mathcal{O}_{0}^{J/\psi}({}^3S_1^{[1]})
|0\rangle$ to obtain the matrix element of order $v^4$ from the matrix
element of order $v^0$ that appears in Ref.~\cite{Butenschoen:2010rq}.
We use the value of $m^2\langle v^2 \rangle$ from Table I of
Ref.~\cite{Bodwin:2007fz}: $m^2\langle v^2 \rangle=0.437\,\textrm{GeV}^2$
at $m=1.5\,\textrm{GeV}$. The results of this computation are shown in
Table~\ref{tab:v4-frag}.

We see from the second row of Table~\ref{tab:v4-frag} that the $Q\bar
Q({}^3S_1^{[1]})$ channel makes a small contribution to $D_4[g\to
J/\psi]$ at ${p_{}}_T=20$~GeV, confirming that this
channel is not important phenomenologically at the current level of
precision. We also see that the $Q\bar Q({}^3S_1^{[8]})$ and $Q\bar
Q({}^3P^{[8]})$ channels give comparable contributions to $D_4[g\to
J/\psi]$ at ${p_{}}_T=20$~GeV. 
We estimate that the fragmentation contribution to
$d\sigma/d{p_{}}_T\times B(J/\psi\to \mu\mu)$ at ${p_{}}_T=20$~GeV from
the $Q\bar Q({}^3P^{[8]})$ channel is about $6\times
10^{-3}$~nb/GeV \cite{Bodwin:2012xc}, which is comparable to 
(about a factor of $2$ larger than) the total NLO contribution 
in the $Q\bar Q({}^3P^{[8]})$ channel in Fig.~1(c) of 
Ref.~\cite{Butenschoen:2010rq}.
A more precise calculation of the fragmentation contribution in the $Q\bar
Q({}^3P^{[8]})$ channel will be necessary in order to determine whether
it is the dominant contribution in that channel at NLO in $\alpha_s$ at
high $p_{_T}$.

\begin{table}
\caption[]{
Relative contributions to $d\sigma_{J/\psi}^{\rm frag}$ in order $v^4$.
The first and second rows give 
$I_{5.2}(d)$ times the LDMEs that are given in the text. We take 
$m=m_c=1.5\,\textrm{GeV}$. For compatibility with 
Ref.~\cite{Butenschoen:2010rq}, we take $\alpha_s=\alpha_s({{m}_{}}_{T})$,
where
${{m}_{}}_{T}=\sqrt{{{p}_{}}_{T}^2+4m_c^2}$. We choose the point
${{p}_{}}_{T}=20$~GeV, which implies that $\alpha_s({{m}_{}}_{T})=0.154$.
\label{tab:v4-frag}}
\begin{center}
\begin{tabular}{llclll}
\hline
\hline
$I_\kappa(d)\,\backslash \,$channel&
$\!\!\!\!\mathcal{O}_{0}^{J/\psi}({}^1S_0^{[8]})$&~&
$\!\!\!\!\mathcal{O}_{0}^{J/\psi}({}^3S_1^{[8]})$&
$\mathcal{O}_{0}^{J/\psi}({}^3P^{[8]})$&
$\mathcal{O}_{4}^{J/\psi}({}^3S_1^{[1]})$
\\
\hline
$I_{5.2}(d)|_{\mu_\Lambda=m\phantom{2}}\times\,10^6\,\textrm{LDME}$&
$\phantom{12}1.65$&&
$\phantom{59}18.6$&
$\phantom{-1}32.7$&
$\phantom{-}0.192$
\\
$I_{5.2}(d)|_{\mu_\Lambda=2m}\times\,10^6\,\textrm{LDME}$&
$\phantom{12}1.65$&&
$\phantom{59}18.6$&
$\phantom{-1}43.6$&
$\phantom{-}0.455$
\\
\hline
\hline
\end{tabular}
\end{center}
\end{table}

\section{Summary\label{sec:discussion}}
We have computed NRQCD short-distance coefficients for gluon
fragmentation into a $J/\psi$ through relative order $v^4$. Our main
result is the expression in Eq.~(\ref{d412sum-ans}) for $d_{4,1}[g \to
Q\bar{Q}({}^3S_1^{[1]})]^{(3)} +d_{4,2}[g \to
Q\bar{Q}({}^3S_1^{[1]})]^{(3)}$, which is the sum of short-distance
coefficients for the ${}^3S_1^{[1]}$ NRQCD LDMEs of relative order $v^4$.
This is the first time that double soft divergences, as well as single
soft divergences, have appeared in intermediate steps in the calculation
of NRQCD short-distance coefficients. We have also computed the free
$Q\bar Q $ NRQCD LDMEs that appear when one uses the matching equations
between full QCD and NRQCD to compute the NRQCD short-distance
coefficients. Our perturbative calculations of these LDMEs involve both
one-loop and two-loop corrections. The ultraviolet
renormalization-subtractions for these LDMEs include Born and one-loop
single-pole counterterm contributions  and Born double-pole counterterm
contributions. We have worked out the renormalization-group evolution of
the renormalized LDMEs and confirm some previous results in
Refs.~\cite{BBL,Gremm:1997dq,z-g-he}, but disagree with a result in
Ref.~\cite{Gremm:1997dq}.

We have estimated the relative sizes of the contributions of the various
channels to gluon fragmentation into a $J/\psi$ through relative
order $v^4$. The contribution in order $v^4$ of the ${}^3S_1^{[1]}$
channel to the cross section at $p_{_T}=20$~GeV  is about a factor of
$2$~$(4)$ larger than the contribution in order $v^0$ when
$\mu_\Lambda=m_c$~$(2m_c)$. In spite of this large enhancement of the
fragmentation contribution in order $v^4$, the corresponding
contribution to the $J/\psi$ production cross section at the Tevatron or
the LHC is not important at the current level of precision.

In the ${}^3S_1^{[8]}$ channel, most of the contribution to the
$J/\psi$ production cross section at hadron-hadron colliders at large
$p_{_T}$ arises from the fragmentation contribution. This is true at LO
in $\alpha_s$ ~\cite{Cho:1995vh} and at NLO in
$\alpha_s$~\cite{Gong:2008ft}. Our estimate of the fragmentation
contribution to the ${}^3P_J^{[8]}$ channel indicates that it is an
important part of the contribution in that channel to the high-$p_{_T}$
$J/\psi$ production cross section at hadron-hadron colliders at NLO in
$\alpha_s$. However, it will be necessary to carry out a more precise
calculation of the fragmentation contribution in the ${}^3P_J^{[8]}$ 
channel in order to see whether it is actually dominant at high $p_{_T}$.
\acknowledgments
The work of G.T.B.~in the High Energy Physics Division at Argonne
National Laboratory was supported by the U.~S.~Department of Energy,
Division of High Energy Physics, under Contract No.~DE-AC02-06CH11357.

\end{document}